\documentclass[fleqn]{an}
\usepackage{amsmath,amssymb}
\usepackage{graphicx, times}
\usepackage{natbib}
\bibpunct{(}{)}{;}{a}{}{,}

\renewcommand{\th}{\vec{\theta}}
\newcommand{\be}{\begin{equation}}
\newcommand{\ee}{\end{equation}}

\newcommand{\M}{\mathcal{M}}
\Pagespan{832}{}
\Yearpublication{2010}%
\Yearsubmission{2010}%
\Month{11}%
\Volume{331}%
\Issue{8}%
\DOI{10.1002/asna.201011409}
\begin{document}

\title{Slicing cluster mass functions with a Bayesian razor}

\author{C.D. Sealfon\thanks{Corresponding author:
  {csealfon@wcupa.edu}}}

\institute{
Department of Physics, West Chester University of Pennsylvania, 
West Chester, Pennsylvania 19383, USA}

\received{2010 Jul 9}
\accepted{2010 Aug 11}
\publonline{later}


\abstract{%
We apply a Bayesian ``razor" to forecast Bayes factors between different parameterizations of the galaxy cluster mass function.
To demonstrate this approach, we calculate the minimum size N-body simulation needed for strong evidence favoring a two-parameter mass function over one-parameter mass functions and visa versa, as a function of the minimum cluster mass.
}

\maketitle

\section{Introduction}

The galaxy cluster mass function describes the abundance of virialized astronomical objects as a function of their 
mass.  The mass function is exponentially sensitive to the initial conditions,
composition, and evolution history of the Universe.  Thus, it can provide a
powerful observational tool to look for features in the standard $\Lambda$CDM
cosmological model, including evolution of the dark energy equation of state
(e.g. \citealt{Voit05, Bhatta10} and references therein) and primordial
non-Gaussianity (\citealt{Desjac10} and references therein).   Numerous
different parameterizations of the mass function have been proposed, with
different numbers of free parameters (see Sect. \ref{sec:massfn}).  Current and future experiments  will observe thousands of galaxy clusters \citep{ACT10, ACT07, SPT08, Planck, Rozo09}, 
producing galaxy cluster catalogs which can be used to constrain the shape of the cluster mass function.  
Yet will this data be sufficient to distinguish statistically significant  differences between parameterizations? 
The minimum number of parameters required by data and/or simulations affects both data analysis (models with fewer parameters are easier to work with) and theory (understanding the physical interpretation of additional parameters as well as their implications for cosmology).   Uncertainties in the mass function parameterization can significantly affect cosmological constraints from cluster abundances \citep*{Cunha10, Wu10}.  

Bayesian model selection is well-suited to address the question of how many mass
function parameters are  re\-quired by data.  Occam's razor implies
that if two theories describe data equally well, the simpler explanation is
preferable.   Thus, to describe a relationship between two physical quantities,
one would like to use a function with the fewest number of parameters necessary.
 Since any physical data (or N-body simulation) has stochastic error, adding
 additional parameters to the function will, in general, decrease the error
 between the function and the data.  This does not mean the data support
 adoption of the extra parameters; in the extreme case, if the function's number
 of degrees of freedom equals that of the data, the error will be zero.  Thus
 one must be careful in comparing models with different numbers of parameters.  
  Bayesian evidence and Bayes factors provide a rigorous way to quantify the
  trade-off between fewer parameters and smaller error (\citealt{Trotta08} and
  references therein), penalizing models with extra parameters unless they fit
  the data significantly better.  It is useful to forecast Bayesian evidence in
  advance of data (e.g.  \citealt*{Heavens07, Trotta07, Muk05}) to determine how well experiments will be able to rule out models.

In this work, we show how to use a ``razor" based on the Kullback-Leibler distance \citep{Vijay1,Vijay2} to forecast the Bayes factors between different models of the galaxy cluster mass function.  
Rather than forecasting the evidence for particular surveys, we raise the more general question of how much data would be required to distinguish among different models of the cluster mass function. 
A mass function model may be written as a probability distribution function with both a continuous part above the minimum discernible cluster mass, and a discrete part  (or a constant probability density) below the cluster mass limit.   As an illustrative example, we examine only the simplest, most ideal case, ignoring sample variance, evolution effects, and measurement errors, such as in a large N-body simulation with well-defined cluster masses at constant redshift.  We demonstrate how the ability to distinguish models depends on the cluster mass limit.  This approach may be extended to estimate the minimum size cluster surveys required to justify or discount additional parameters in the cluster mass function.

  In Sect. \ref{sec:massfn} we give an overview of cluster mass function models,
  and in Sect. \ref{sec:razor} we define the razor and demonstrate its
  application to models with mixed probability distribution functions.  We then
  apply the razor in Sect. \ref{sec:mfap} to distinguish between a
  two-parameter  mass function and one-parameter mass functions.  We conclude in
  Sect. \ref{sec:conc}.

\section{Cluster mass function}
\label{sec:massfn}

The mass function $n(m,z) dm$ is the comoving number density 
of galaxy clusters with mass between $m$ and $m+dm$ at redshift $z$.
As in \citet{ST99}, we write the mass function in terms of the dimensionless parameter 
$\nu \equiv \frac{\delta_c^2}{\sigma^2(m,z)}$, where $\delta_c$ is 
the critical overdensity for collapse, $\sigma^2(m,z)$ is the variance of density fluctuations smoothed on scales 
$r= (3 m / 4 \pi \bar{\rho} )^{1/3}$, and $\bar{\rho}$ is the mean matter density of the universe. 
The mass function is then given by the dimensionless function $f(\nu)$, where
\be
f(\nu) d \nu = \frac{m}{\bar{\rho}} n(m,z) \, dm .
\ee
The mass function thus also gives the probability density $p(\nu)$ that a particle with mass $\delta m \ll m$
is in a cluster with a mass parameter between $\nu$ and $\nu + d \nu$ \citep*{Manera09}: 
$
p(\nu) d \nu =  f(\nu) d \nu .
$
 
Since the formation of galaxy clusters is a highly nonlinear problem, the exact shape of the cluster mass function in $\Lambda$CDM and alternative cosmologies remains an open question of great interest and active research \citep{Bhatta10}.  
Analytical models for the cluster mass function have been developed using the
excursion set approach (\citealt*{PS74, Bond91, ST01, Zentner07} and references therein), which differ from fitting functions based on numerical simulations presented by, e.g.,  \citealt{Jenkins01, Evrard02, Reed06, Warren06, Tinker08, Crocce10, Bhatta10}.   The recent fitting functions introduce extra parameters which do not currently have an underlying physical interpretation  \citep{Robertson09}.  We demonstrate a new application of Bayesian model selection which can quantify the statistical significance of the differences among different functional forms of the cluster mass function.

\section{Bayesian razor for a mixed probability distribution}
\label{sec:razor}

Within a Bayesian statistical framework, the Bayesian evidence of different models may be used to compare their relative statistical significance given certain data \citep{Trotta08}.  For a model $\mathcal{M}$ with $n$ parameters 
$\vec{\theta}=\{\theta_1, \theta_2, ..., \theta_n\}$ and a data set of $N$ outcomes $\vec{ \nu} = \{\nu_1, \nu_2, ..., \nu_N\}$ 
drawn independently from a fiducial underlying probability density function $ p_\mathrm{true}({\nu})$, the evidence is
\be
p(\vec{\nu}|\mathcal{M}) \equiv \int d^n \theta \, \Pi(\vec{\theta}| \mathcal{M}) \,  p(\vec{\nu}|\vec{\theta}, \mathcal{M})
\label{eq:ev}
\ee
where $\Pi(\vec{\theta}| \mathcal{M})$ is a normalized prior distribution and  $p(\vec{\nu}|\vec{\theta}, \mathcal{M})$ is the likelihood.

 The expectation of the log likelihood, $\langle \ln p(\vec{\nu}|\vec{\theta}, \mathcal{M}) \rangle$, is equal to the Kullback-Leibler distance  between the true distribution and the model distribution,  $D( p_\mathrm{true}({\vec{\nu}}) || p(\vec{\nu}|\vec{\theta}, \mathcal{M}))$, plus a constant that depends only on the entropy of the true distribution.  Following \citet{Vijay1, Vijay2}, 
 we define the razor of a model, $R(\mathcal{M})$:
\be
R(\mathcal{M}) \equiv  \int{ d^n \theta \,  \Pi(\vec{\theta}| \mathcal{M}) \,  e^{-N D( p_\mathrm{true}({\nu}) || p({\nu}|\vec{\theta}, \mathcal{M}))}}.
\ee
The ratio of the razors of two models thus may be used to forecast the Bayes factor (the ratio of the evidences) for a given fiducial model.

Suppose we can only distinguish data with values of $\nu$ above a certain limit $\nu_d$.  (In the context of the mass function, $\nu_d$ will correspond to the minimum cluster mass, known as the dust limit.)  Define $f_d$ to be the fraction of the outcomes with $\nu<\nu_d$, that is
\be
f_d =  \int_0^{\nu_d} d \nu \, p(\nu) .
\ee
This situation results in  a mixed probability distribution  function, with a discrete ``bin" for the fraction of data with $0<\nu<\nu_d$ (with a constant probability density $\frac{f_d}{\nu_d}$) and a continuous probability distribution for data with $\nu>\nu_d$.  The Kullback-Leibler (KL) distance is in this case:
\begin{eqnarray}
D(\th, \M) &\equiv& D( p_\mathrm{true}({\nu}) || p({\nu}|\vec{\theta}, \mathcal{M})) \nonumber\\
&=&{ f_d}_\mathrm{true} \ln \left( \frac{ { f_d}_\mathrm{true} }  { f_d (\vec{\theta},\M) }  \right)  \\
&& + \int_{\nu_d}^\infty  d \nu  \, p_\mathrm{true}({\nu}) \ln \left( \frac{  p_\mathrm{true}({\nu}) }  { p({\nu}|\vec{\theta}, \mathcal{M}) }  \right) . \nonumber
\end{eqnarray}

In the Laplace approximation for large $N$ \citep{MacKay, Trotta08}, we can Taylor expand the KL distance around its minimum, keeping the first two terms:
\begin{equation}
R(\mathcal{M}) \! \simeq \! \! \int \! \! \! d^n \delta \,  \Pi(\vec{\theta}| \mathcal{M}) \, 
e^{\! \! -N \left(\!D(\vec{\theta}_0 ,\M)    + \frac{1}{2} \delta_i \delta_j \frac{\partial^2}{\partial \theta_i \partial \theta_j} D (\vec{\theta}_0,\M )\!\right)}
\end{equation}
where $\vec{\delta} = \vec{\theta_0} - \vec{\theta}$, and $\th_0$ represents the values of the parameters that minimize the KL distance with a given fiducial model.
Assuming a flat prior, $\Pi(\vec{\theta}| \mathcal{M})$ is given by the reciprocal of the volume of the parameter space, $\Pi(\vec{\theta}| \mathcal{M}) = (\Delta \theta_1 \Delta \theta_2 ... \Delta \theta_n)^{-1}$.  For large $N$, the likelihood outside the boundaries of the parameter space is negligible, and we can estimate the razor integral as the integral over the entire Gaussian.  Noticing that $\frac{\partial^2}{\partial \theta_i \partial \theta_j} D (\vec{\theta}_0,\M )$ equals the Fisher matrix $F_{ij}(\th_0,\M)$
, we can write

\begin{equation}
R(\mathcal{M}) \simeq \Pi(\vec{\theta}| \mathcal{M})  e^{-N D(\vec{\theta}_0 ,\M) } 
\sqrt { \frac{(2 \pi)^n}{N^n \det(F_{ij}(\th_0,\M))}} .
\end{equation}
   The log razor (for a flat prior) is
\begin{eqnarray}
\label{eq:razor}
\ln R(\M) &\simeq  & - \ln (\Delta \theta_1 ... \Delta \theta_n)  +\frac{n}{2} \ln 2 \pi  \\
&& -\frac{1}{2} \ln  \det(F_{ij}(\th_0,\M))  
  -\frac{n}{2} \ln N \nonumber \\ && 
-N D(\vec{\theta}_0 ,\M). \nonumber
\end{eqnarray}

To compare two models $\M_1$ and $\M_2$, we take the ratio of their razors.  Suppose model $\M_1$, with $m$ parameters, is nested inside model $\M_2$, with $n>m$ parameters.  Then the log razor ratio reduces to
\begin{eqnarray}
\label{eq:ratio}
\ln \frac{R(\M_1)}{R(\M_2)} &\approx& \ln (\Delta \theta_{m+1} ... \Delta \theta_n) - \frac{n-m}{2} \ln (2 \pi)  \nonumber \\
&& + \frac{1}{2} \ln \frac{ |F_{ij}(\th_{02},\M_2)|}{| F_{ij}({\th}_{01},\M_1)|}  \\
&& +\left( \frac{n-m}{2} \right)\ln N - N \left(D_{01}-D_{02} \right) \nonumber
\end{eqnarray}
where $D_{01}\equiv D(\vec{\theta_0}_1 ,\M_1)$ and $D_{02} \equiv
D(\vec{\theta_0}_2 ,\M_2) $ are the minimum KL distances for each
model.  This expression is equivalent to the log of Eq. (10) in \citet{Heavens07}.
Note that a positive log razor ratio favors model $\M_1$ and a negative log razor ratio favors model $\M_2$.
The KL distance equals zero when the model distribution equals the true distribution.  If the more complicated model $\M_2$ is true, then $D_{01}>D_{02}=0$, and the more complicated model will be favored for large $N$ with a log razor ratio decreasing linearly with $N$.  If the simpler model $\M_1$ is true, then $D_{01}=D_{02}=0$ (since $\M_1$ is nested inside $\M_2$).  In this case, for large $N$, the simpler model will be favored with a log razor ratio proportional to $\ln (N)$.  

The razor, like Bayesian evidence, effectively measures how much of the prior volume in parameter space is taken up by the posterior distribution.  For flat priors and Gaussian likelihoods, in the limit of large $N$, the integral of the posterior over the parameter space will be practically independent of the boundaries of the parameter space.  However, the parameter-space volume, $\Delta \theta_1 ... \Delta \theta_n$, depends sensitively on these bounds; the greater this volume, the smaller the razor.  Thus the razor ratio will depend on the choice of parameter ranges or prior distribution considered.  
Any broad, smoothly varying prior may be approximated as a constant near the peak of the likelihood term (that is, the peak of $e^{-N D(\vec{\theta}_0 ,\M) }$) in the limit of large $N$, and thus may be estimated in the Laplace approximation by a flat prior with value $\Pi(\vec{\theta}_0| \mathcal{M})$ (or an effective  parameter-space volume of $\frac{1}{\Pi(\vec{\theta}_0| \mathcal{M})}$).

The razor also depends on how narrowly peaked the posterior is, through $\det( F_{ij})$.  A greater value of $\det (F_{ij})$ means greater curvature at the maximum of the posterior (a narrower peak) and thus a smaller volume under the posterior.  So if two models have the same prior parameter-space volume and the same maximum likelihood, the model with the larger value of $\det (F_{ij})$ will have a smaller razor.  This is slightly counter-intuitive because one might naively expect a ``narrower" model to be ``simpler" and thus favored by Occam's razor.  However, the razor is concerned with the ratio of posterior volume to prior parameter-space volume, and the ``narrower" model wastes more of the available parameter space.

In the limit of large $N$, the log razor in Eq. (\ref{eq:razor}) approaches the expectation of  the Bayesian Information Criterion (BIC) times negative one-half (cf. \citealt{Trotta08}, Eq. 37).\footnote{Note that the popular Akaike Information Criterion (AIC) is based on a fundamentally different model-comparison approach, which estimates the expected error in the maximum log likelihood for each model (cf. \citealt{Takeuchi99, Liddle09}).}

\section{Applying the razor to the cluster mass function}
\label{sec:mfap}

The razor  allows  comparison of different functional forms of the cluster mass function $f(\nu)$.  
To demonstrate, we compare the two-parameter Sheth-Tormen (ST) mass function \citep{ST99},

\be
f(\nu) \propto \sqrt{\frac{a}{2 \pi \nu}} \left(1+ (a \nu)^{-p} \right) 
e^{\frac{-a \nu}{2} }, 
\label{eq:ST}
\ee
with one-parameter mass functions keeping $a$ or $p$ constant:
\be
f(\nu) = \sqrt{\frac{a}{2 \pi \nu}} 
e^{\frac{-a \nu}{2} }
\label{eq:PSa}
\ee
(with $p=0$) or
\be
f(\nu) \propto \sqrt{\frac{1}{2 \pi \nu}} \left(1+ \nu^{-p} \right) 
e^{\frac{-\nu}{2} }
\label{eq:PSp}
\ee
(with $a=1$).  Note that the Press-Schechter (PS) mass function \citep{PS74} corresponds to $a=1$ and $p=0$.

\begin{figure*}[!htp]
\begin{tabular}{cc}
\includegraphics[width=8 cm]{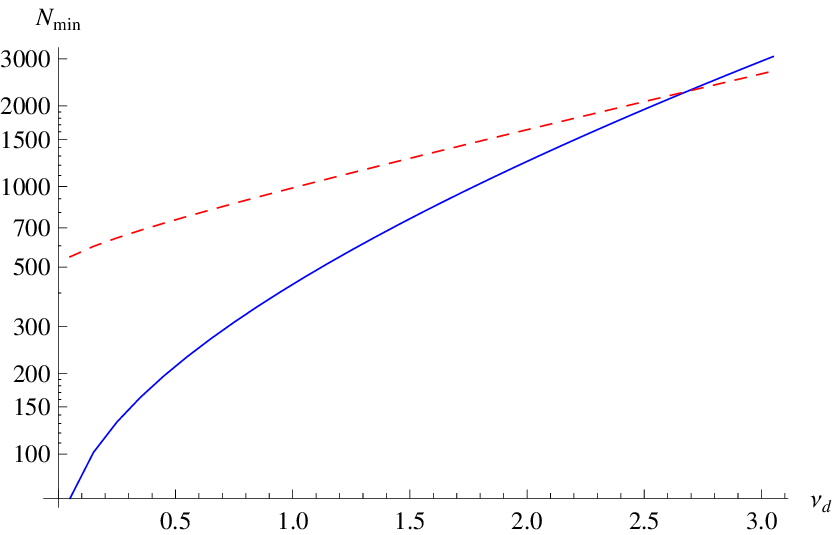} &
\includegraphics[width=8 cm]{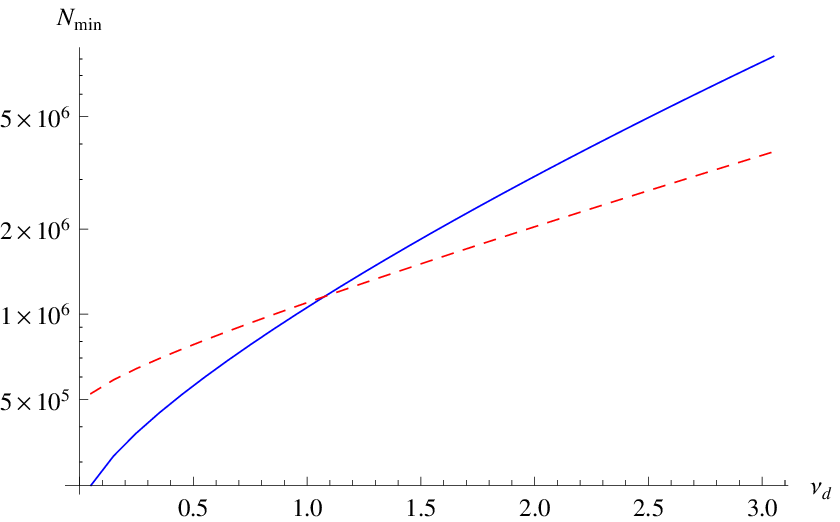}
\end{tabular}
\caption{$N_\mathrm{min}$ vs. $\nu_d$ where model $\M_2$ is the 2-parameter ST model.  For the blue solid line, model $\M_1$ is the ST model with only $a$ as a free parameter ($p$ is fixed at $p=0$).   For the red dashed line,  model $\M_1$ is the ST model with only $p$ as a free parameter ($a$ is fixed at $a=1$).  {\it Left:} The fiducial model is ST with $a=0.7$ and $p=0.3$ (requiring both parameters). {\it Right:} The fiducial model is the Press-Schechter model ($a=1,p=0$).}
\label{fig:v}
\end{figure*}

\begin{figure*}[!htp]
\begin{tabular}{cc}
\includegraphics[width=8 cm]{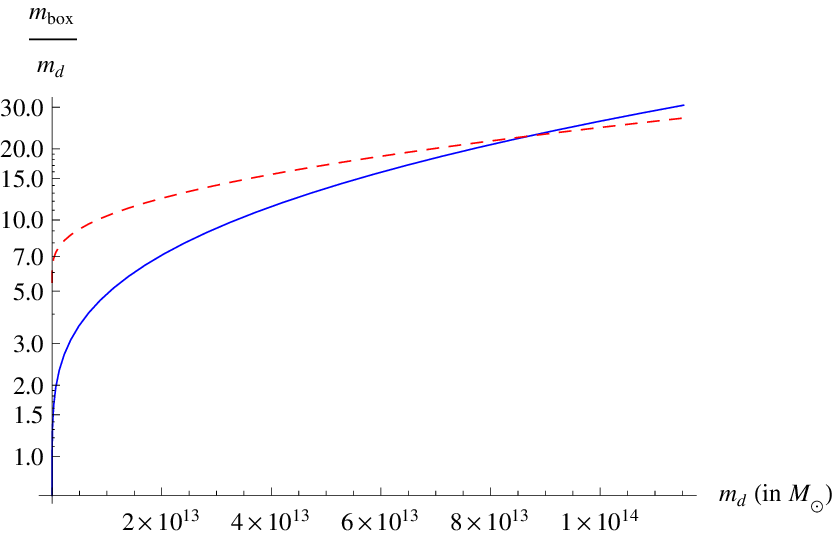} &
\includegraphics[width=8 cm]{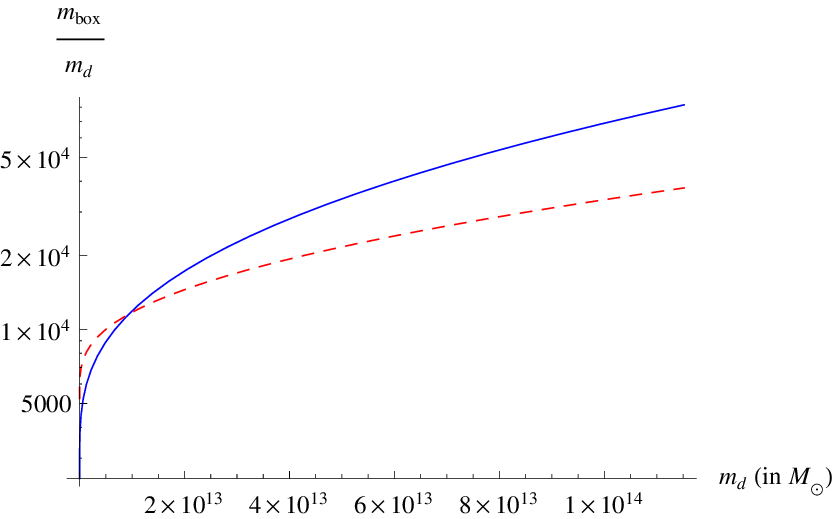}
\end{tabular}
\caption{$\frac{m_\mathrm{box}}{m_d}$ vs. $m_d$ in solar masses, where model $\M_2$ is the 2-parameter ST model.  For the blue solid line, model $\M_1$ is the ST model with only $a$ as a free parameter ($p$ is fixed at $p=0$).   For the red dashed line,  model $\M_1$ is the ST model with only $p$ as a free parameter ($a$ is fixed at $a=1$). {\it Left:} The fiducial model is ST with $a=0.7$ and $p=0.3$ (requiring both parameters). {\it Right:} The fiducial model is the Press-Schechter model ($a=1,p=0$).}
\label{fig:mass}
\end{figure*}

We seek to estimate $N_\mathrm{min}$, the minimum number of particles (data)
needed for strong evidence in favor of the ``true" or fiducial model.  On a
Jeffreys scale (e.g.  \citealt{Trotta08}), strong evidence corresponds to
$\left|\ln \frac{R(\M_1)}{R(\M_2)}\right| \gtrsim 5$.  From Eq. (\ref{eq:ratio}),  $N_\mathrm{min}$ will clearly depend on the chosen prior volume for the extra parameters in $\M_2$.  If the simpler model $\M_1$ is the true model,   $N_\mathrm{min}$ will scale as  $N_\mathrm{min} \propto \left(\Delta \theta_{m+1} ... \Delta \theta_n \right)^{\frac{-2}{n-m}}$.  

We  calculate how $N_\mathrm{min}$ depends on the dust limit $\nu_d$, assuming
parameter ranges of $\Delta a = 1$ and $\Delta p = 1$.  Figure~\ref{fig:v} shows
$N_\mathrm{min}$ versus $\nu_d$ needed to distinguish between the ST model and
the one-parameter models: the left plot for a fiducial ST model with $a=0.7$ and
$p=0.3$, and the right plot for a PS fiducial model.  The blue solid lines
correspond to the one-parameter model with $p=0$ (Eq. (\ref{eq:PSa})), and
the red dashed lines, the one-parameter model with $a=1$ (Eq. (\ref{eq:PSp})).

For the case where the two-parameter ST model is the fiducial model
(Fig.~\ref{fig:v} left), the razor ratio will be negative and dominated by the
KL distance between the simpler model and the fiducial model ($- N D_{01}$) for
large $N$.    Equation (\ref{eq:PSa}) can more closely approximate the fiducial
ST model at large $\nu$, but Eq. (\ref{eq:PSp}) can more 
closely approximate the fiducial model at small $\nu$.  We can thus see why, for
small $\nu_d$, it takes more particles to distinguish the fiducial model from
$\M_1$ given by Eq. (\ref{eq:PSp}) than Eq.  (\ref{eq:PSa}): the value of
$D_{01}$ is smaller for model Eq.  (\ref{eq:PSp}).  For large values of $\nu_d$,
the situation is reversed, and model Eq. (\ref{eq:PSa}) is harder to rule out
than model Eq. (\ref{eq:PSp}).

For the case where the PS model is true, the minimum KL distance between the
fiducial model and all the parametric models is zero, and the razor ratio is
positive and dominated by the $\ln N$ term for large $N$.  The ability to
distinguish the simpler models depends on the value of  $|
F_{ij}({\th}_{01},\M_1)|$, that is, the curvature of the log likelihood near its
maximum; the greater the curvature, the lower the evidence of the simpler model
compared to the ST model, and the more particles are needed to prefer the
simpler  model.  Model Eq.  (\ref{eq:PSa}) $(p=0)$ deviates from PS
more quickly at high $\nu$, and thus has a higher value of  $|
F_{ij}({\th}_{01},\M_1)|$ than model Eq.  (\ref{eq:PSp}) $(a=1)$ for high values of $\nu_d$.  

We can relate the dust limit $\nu_d$ to a mass limit $m_d$ by assuming a
standard $\Lambda$CDM cosmology, approximating the power spectrum using the
fitting function given by Eq. (7) in  \citet*{Efs92}  (see also \citealt{Bond84}) with $h=0.71$ and $\Omega_m = 0.27$,  and taking $\delta_c=1.69$.  
We assume that a minimum of 100 particles form a cluster, and set the mass of a particle to be $\delta m = m_d/100$.  Then we  convert the minimum number of particles $N_\mathrm{min}$ to the mass of these particles, $m_\mathrm{box}$, via $m_\mathrm{box}= \frac{N_\mathrm{min} m_d }{ 100}$.  For $m_\mathrm{box} \gg m_d$,  $m_\mathrm{box}$ gives the minimum size of a simulation box needed to distinguish among models, given a dust limit and a fiducial underlying model.
\footnote{
Our analysis assumes the $N$ particles are drawn from clusters of all possible masses, using the likelihood function from \cite{Manera09} Appendix A.  When $\frac{m_\mathrm{box}}{m_d} \sim$ a few, however, there are clearly just a few clusters in the simulation box, and in that case,  $m_\mathrm{box}$ provides a lower limit on the simulation size actually needed to distinguish models.  
} 
These results, effectively just a transformation of the graphs in Fig.
\ref{fig:v},  are shown in Fig. \ref{fig:mass}.  Clearly, the size of the
simulation needed to distinguish models increases with the minimum cluster mass
$m_d$.  We see that a larger (smaller) number of particles in
Fig. \ref{fig:v} corresponds to a larger (smaller) ratio of the simulation mass to the minimum cluster mass required to distinguish models.

\section{Discussion}
\label{sec:conc}

We have demonstrated a new application of the Bayesian razor to estimate the necessary N-body simulation size to distinguish among different models of the cluster mass function, with different numbers of free parameters.  Our approach quantifies how this simulation volume depends on the minimum cluster mass $m_d$.  Reducing the dust limit significantly enhances the ability to distinguish models, as the mass of the simulation in units of $m_d$ increases by roughly an order of magnitude as $m_d$ increases from $10^{12} M_\odot$ to $10^{14} M_\odot$.  In general, it is much more difficult to have strong evidence against a complicated model than to strongly favor it.  This is because the log razor ratio goes as $\ln N$ when the simpler model is true, but goes as $N$ when the more complicated model is true.  In our  examples, simulations must be thousands of times larger to favor the true model for fiducial one-parameter models than for the fiducial two-parameter model.  
Future work will extend our analysis to compare higher-dimensional mass function parameterizations and to incorporate redshift-evolution effects, sample variance, and measurement errors for cluster surveys.  
We note that the application of the razor to a mixed probability distribution may also be used in other applications of Bayesian model comparison where certain ranges of data are binned.

\acknowledgements{The author gratefully acknowledges Ravi Sheth for discussions that stimulated this investigation and  Kavilan Moodley for helpful comments to improve the manuscript. Sincere thanks to them, Darell Moodley, Benjamin Burner, and Devin Crichton for helpful discussions.
}

\bibliographystyle{apj}

\end{document}